# Enhanced photoresponse in MoTe$_2$ photodetectors with asymmetric graphene contacts

*Xia Wei, Faguang Yan, Quanshan Lv, Wenkai Zhu, Ce Hu, Amalia Patanè and Kaiyou Wang*[*]

Mr. X. Wei, Dr. F. Yan, Mr. Q. Lv, Mr. W. Zhu, Mr. C. Hu and Prof. K. Wang
State Key Laboratory of Superlattices and Microstructures, Institute of Semiconductors, Chinese Academy of Sciences, Beijing 100083, P. R. China
Center of Materials Science and Optoelectronics Engineering, University of Chinese Academy of Sciences, Beijing 100049, P. R. China
Center for Excellence in Topological Quantum Computation, University of Chinese Academy of Science, Beijing 100049, P. R. China
E-mail: kywang@semi.ac.cn
Prof. A. Patanè
State Key Laboratory of Superlattices and Microstructures, Institute of Semiconductors, Chinese Academy of Sciences, Beijing 100083, China
School of Physics and Astronomy, University of Nottingham, Nottingham NG7 2RD, United Kingdom

**Keywords:** asymmetric graphene contacts, MoTe$_2$, van der Waals heterostructure, built-in electric filed, enhanced photoresponse

**Atomically thin two dimensional (2D) materials are promising candidates for miniaturized high-performance optoelectronic devices. Here, we report on multilayer MoTe$_2$ photodetectors contacted with asymmetric electrodes based on *n*- and *p*-type graphene layers. The asymmetry in the graphene contacts creates a large ($E_{bi}$ ~100 kV cm$^{-1}$) built-in electric field across the short ($l$ = 15 nm) MoTe$_2$ channel, causing a high and broad ($\lambda$ = 400 to 1400 nm) photoresponse even without any externally applied voltage. Spatially resolved photovoltage maps reveal an enhanced photoresponse and larger built-in electric field in regions of the MoTe$_2$ layer between the two graphene contacts. Furthermore, a fast (~10 μs) photoresponse is achieved in both the photovoltaic and photoconductive operation modes of the junction. Our findings could be extended to other 2D materials and offer prospects for the implementation of asymmetric graphene contacts in future low-power optoelectronic applications.**

# 1. Introduction

Two dimensional (2D) van der Waals crystals have received great attention due to their excellent properties and versatility for a wide range of potential applications in optoelectronics.[1,2] In particular, transition metal dichalogenides (TMDs) with their finite and tunable bandgap energy (from $E_g$ = 1.1 to 2.1 eV) and strong light absorption offer opportunities for a variety of optoelectronic devices.[2-4] Amongst the TMDs, MoTe$_2$ is an attractive semiconductor. In the monolayer form, it has a direct bandgap, $E_g$ = 1.10 eV at room temperature, larger than that of bulk MoTe$_2$, which has an indirect bandgap ($E_g$ = 0.85 eV).[5-7] Thus unlike other TMDs, such as MoS$_2$ and WS$_2$, photodetectors based on MoTe$_2$ can have a broadband photoresponse that extends from the visible (VIS) to the near infrared (NIR) spectral range.[8-10] In particular, in MoTe$_2$-based field effect transistors (FETs), the photoresponsivity ($R$) can be enhanced by a photogating effect and achieve values of up to $R$ = 24 mA W$^{-1}$ under illumination with NIR light.[9]

In contrast to traditional bulk semiconductors such as Si or III-V compounds, 2D vdW crystals have pristine surfaces that are free of dangling bonds. This offers opportunities to combine them with other materials without the limitations of lattice mismatch that apply to covalent crystals.[2,11] For example, MoTe$_2$ has been used in different multilayer structures: in MoTe$_2$/MoS$_2$ heterojunctions, the on/off photocurrent ratio can reach values of about 780;[12] also, the photoconductive gain in MoTe$_2$/graphene heterostructures can be as large as 4.69×10$^8$.[9] More generally, asymmetric contact barriers between two electrodes and a 2D vdW crystal can be exploited to construct high performance photodetectors:[13-20] Au and In Schottky contacts to a 2D material can be used to realize self-powered photodetectors with high photoresponsivity ($R$ = 110 mA W$^{-1}$).[13] Also, graphene can form a clean interface with 2D materials and its near perfect optical transparency makes it suitable for use as the top electrode of vertical heterostructure photodetectors.[21-24] Au/MoTe$_2$/graphene vertical heterostructures have good photoresponsivity and photoreponse time of about 96 ms.[14] However, the photoresponse of 2D vdW heterostructure devices in the current literature remain still slow due to relatively long optically active channels and/or charge traps at the metal/2D material interface. Thus, both the length of the channel and the quality of the contacts should be carefully chosen to optimize the photoresponse.

In this study, we report on photodetectors based on MoTe$_2$ with vertical asymmetric graphene contacts. We use *p*-type graphene grown by chemical vapour deposition (CVD) as the top contact and *n*-type exfoliated graphene as the bottom contact. This asymmetry in the graphene contacts is adopted to break the mirror symmetry of the internal electric field profile, thus creating a large built-in electric field $E_{bi}$. This feature combined with the short length of the MoTe$_2$ optically active channel enables an efficient and fast photoresponse. The heterostructure exhibits a high and broad spectral photoresponse from the VIS to the NIR range of the electromagnetic spectrum ($\lambda$ = 400 – 1400 nm) without any external applied voltage: the photoresponsivity is $R$ = 12.38 mA W$^{-1}$ at $\lambda$ = 1064 nm and $R$ = 27.64 mA W$^{-1}$ at $\lambda$ = 550 nm. Through scanning photovoltaic mapping, an enhanced light absorption is clearly observed in the overlapping region of the graphene and MoTe$_2$ layers. Furthermore, because of the short ($l$ = 15 nm) MoTe$_2$ channel, the response time of the heterostructure can be as short as ~ 6.15 μs, which is 1-3 orders of magnitude faster than that reported before for MoTe$_2$-based photodetectors.[1,9,14,25,26]

## 2. Results and discussion

**Figure 1**(a) shows the schematic layout of our MoTe$_2$-graphene based photodetectors. To fabricate the MoTe$_2$-graphene heterostructure, a multilayer graphene flake was first mechanically exfoliated from a bulk crystal using adhesive tape and then transferred onto a Si/SiO$_2$ substrate (300 nm-thick SiO$_2$).[27,28] Using the same mechanical exfoliation and transfer method, a flake of MoTe$_2$ was then transferred onto the bottom graphene layer. Finally, a microstamp of CVD-graphene was transferred on top of the MoTe$_2$ flake to create the top electrode.[28] Metallic contacts (Ta/Au) were fabricated on the substrate using standard photoetching, magnetron sputtering and lift off. Single crystals of 2H-MoTe$_2$ were purchased from HQ graphene (Netherlands); CVD-grown graphene was provided by G-CVD (Xiamen, China) and the bulk graphite was purchased from 2D Semiconductors (US). All mechanical exfoliation and transfer processes were conducted inside a glove box.

Figure1(b) shows the atomic force microscopy (AFM) image of one of our devices. The large and uniform exfoliated multi-layer graphene serves as a bottom electrode and has a thickness of 3.6 nm; the MoTe$_2$ flake has a thickness of about 15 nm; the top CVD-graphene layer corresponds to single layer graphene. Raman spectroscopy studies were conducted to assess the quality of the individual flakes and heterostructure over an extended frequency range (from 200 to 2800 cm$^{-1}$) with an excitation laser wavelength $\lambda$ = 532 nm and power $P$ = 100 μW. A large magnification (100 ×) objective lens was used to focus the laser to a spot diameter of 1.5 μm and probe different regions, including the MoTe$_2$, exfoliated graphene, CVD-graphene and the region where all layers overlap.

Figure 1(c) shows typical Raman spectra of our samples. The Raman spectrum for the isolated MoTe$_2$ flake shows two Raman-active modes: the $E_{2g}^{1}$ mode at 232.14 cm$^{-1}$ and the $B_{2g}^{1}$ mode at 288.40 cm$^{-1}$, as observed in the literature.[29] For bulk MoTe$_2$, only the $E_{2g}^{1}$ mode is observed (**Figure S1**(a)), consistent with previous reports.[5,7] The difference between the bulk and 2D layers is assigned to the breakup of the translation crystal symmetry in few layer MoTe$_2$.[9] In the Raman spectra of CVD and exfoliated graphene, the most intense features are the G and 2D peaks. The 2D peak from CVD graphene is approximately twice more intense than the G peak, confirming that CVD graphene is a single layer.[30] In contrast, the G peak for exfoliated graphene is higher than the 2D peak, indicating that the flake is multilayer, in agreement with the AFM data. In the Raman spectra of the vertical overlapping region, all the above peaks can be clearly observed, demonstrating the good quality of the heterostructure.

To investigate the electrical properties of our devices, the current, $I_{ds}$, was measured for different voltages, $V_{ds}$, applied between the drain (top CVD graphene) and source (bottom exfoliated graphene) contacts. All the electrical measurements were carried out in vacuum (~5 mbar) at room temperature. Figure 1(d) shows the $I_{ds}$ - $V_{ds}$ curves in the dark and under illumination: in the dark, the $I_{ds}$ - $V_{ds}$ curve shows nonlinear rectifying characteristics, consistent with asymmetric contact barriers between the MoTe$_2$ flakes and the two graphene electrodes. Under continuous-wave laser illumination ($\lambda$ = 1064 nm), the current increases and the $I_{ds}$ - $V_{ds}$ shows a clear photovoltaic effect. For comparison, we fabricated photodetectors based on MoTe$_2$ with symmetric graphene contacts. In this case, the rectification behavior is not observed or is much smaller (inset of Figure 1(d)).

We now examine in detail the photoresponse properties of our photodetectors. **Figure 2**(a)

shows the $I_{ds}$ - $V_{ds}$ characteristics with 1064 nm laser illumination at powers ranging from $P$ = 0 to 6.25 W cm$^{-2}$. The photocurrent, $I_{ph}$, is defined as $I_{ph}=|I_{light}-I_{dark}|$ where $I_{light}$ and $I_{dark}$ are the currents measured with and without illumination, respectively. As shown in Figure 2(b), the photocurrent $I_{ph}$ increases with increasing $P$, revealing a sublinear behavior, i.e. $I_{ph} \propto P^{\alpha}$, where α increases from 0.855 to 0.911 when $V_{ds}$ decreases from 0 to -0.20 V. Thus even without any applied source-drain voltage, the photo-generated electron-hole pairs can be effectively separated by the built-in electric field of the heterostructure to generate a photocurrent. Also, although α increases with decreasing $V_{ds}$, its value remains always smaller than 1. This is the fingerprint of a photoconductivity gain that is influenced by charge traps in the layers and/or their interfaces.[9,31,32] A similar phenomenon was reported in heterostructures based on different vdW crystals, such as WS$_2$/MoS$_2$ and WSe$_2$/GaSe heterojunctions.[33,34]

Figure 2(d) plots the photoresponsivity ($R=I_{ph}/PS$) of the heterostructure at different applied voltages as a function of the incident laser power $P$ and $\lambda$ = 1064 nm. Here $S$ is the in-plane area (~ 400 μm$^2$) of the heterostructure device. The photoresponsivity increases with increasing reverse bias, reaching a value of $R$ = 65.56 mA W$^{-1}$ at $V_{ds}$ = -0.20 V and $P$ = 12.50 mW cm$^{-2}$. The photoresponsivity remains high at $V_{ds}$ = 0 V: $R$ = 12.38 mA W$^{-1}$, which is larger than that reported for MoTe$_2$ layers.[35] The photoresponse retains similar characteristics at different wavelengths (**Figure S2**) with a photoresponsivity of up to $R$ = 27.64 mA W$^{-1}$ at $\lambda$ = 550 nm and $P$ = 12.50 mW cm$^{-2}$. Thus rectification and photovoltaic effects can be clearly observed in MoTe$_2$ layers with vertical asymmetric graphene contacts. The non-zero open-circuit voltage ($V_{oc}$) and short-circuit current ($I_{sc}$) increase with increasing laser power (**Figure S3**(a-b)). For photodetectors based on MoTe$_2$ with symmetric exfoliated graphene contacts, the photoresponse is significantly weaker (**Figure S4**).

**Figure 3**(a) shows the normalized spectral response of the heterostructure at different illumination wavelengths (under the same power $P$ = 0.63 W cm$^{-2}$). It reveals a broad photoresponse from the VIS to the NIR range ($\lambda$ = 400 – 1400 nm) with two clear peaks around $\lambda \sim$ 600 nm and 1150 nm. The peak at $\lambda \sim$ 1150 nm corresponds to the excitonic absorption associated with the lowest direct optical transition at the K-point of the Brillouin zone.[5,36] The peak at $\lambda \sim$ 600 nm arises from high energy excitonic transitions influenced by interlayer interactions.[5,37] A similar peak was also observed in the reflectance spectra of thin MoTe$_2$ layers.[5] The weak photoresponse at long wavelengths ($\lambda \sim$ 1400 nm) is assigned to the optical absorption from band tail states due to charge traps.[8,9]

To investigate the photoresponse in further detail, we carried out a series of measurements of the spatially resolved photovoltage maps. These were obtained by scanning a focused laser beam across the plane of the device without any externally applied voltage. Figure 3(b) shows the optical image of the heterostructure. The corresponding normalized photovoltage maps at $\lambda$ = 550 nm and 1064 nm are shown in Figure 3(c) and (d), respectively. In each Figure, the different parts of the heterostructure are marked in different colors: the bottom exfoliated graphene is marked with a white solid line; the MoTe$_2$ flake is marked with a yellow solid line, and the top CVD graphene is highlighted by a blue solid line. As shown in Figure 3(c) and (d), the photovoltage is non-uniform and is enhanced in the region of the heterostructures where all layers overlap. This enhancement was not observed in previous studies of MoTe$_2$ with symmetric CVD graphene electrodes in the literature.[35] Also, in our samples, regions with exfoliated graphene/MoTe$_2$ and MoTe$_2$/CVD graphene, the photovoltage is much weaker. The measured photovoltage is positive in all regions, implying the existence of a built-in electric field pointing in the same direction, *e.g.* from the bottom exfoliated graphene to MoTe$_2$ in exfoliated graphene/MoTe$_2$, from the bottom exfoliated graphene to CVD graphene in exfoliated graphene/MoTe$_2$/CVD graphene, and from MoTe$_2$ to CVD graphene in MoTe$_2$/CVD graphene.

To estimate the built-in electric field, $E_{bi}$, in the heterostructure, we fabricated and measured the transfer characteristics of FETs based on individual MoTe$_2$, exfoliated graphene, and CVD-graphene layers on a SiO$_2$/Si substrate. According to the transfer curves in **Figure S5**, multi-layer MoTe$_2$ is *p*-type, exfoliated graphene is *n*-type and CVD graphene is *p*-type.

The Fermi energy, $E_F$, for monolayer graphene, is derived from equation (1):[33,38,39]

$$E_F = sign(V_g\text{-}V_D)\hbar|v_F|\sqrt{(\pi\varepsilon\varepsilon_0/te)|V_g\text{-}V_D|} \quad . \tag{1}$$

Here $E_F$ is measured relative to the neutrality point of the graphene band structure, $V_g$ is the back gate voltage on the Si-substrate, $V_D$ is the charge neutrality point voltage from the transfer curves of graphene, $v_F \approx 10^6$ m s$^{-1}$ is the Fermi velocity, $\varepsilon_0$ is the permittivity of free space, $\varepsilon = 3.9$ is the dielectric constant of SiO$_2$, $e$ is the electron charge and $t = 300$ nm is the thickness of SiO$_2$.[33,38,39] The Dirac point is set at -4.55 eV relative to the vacuum level.

From the measured transfer curves in Figure S5, we derive that the Fermi level of CVD-graphene is at $E_{Fp}$ = -4.74 eV at $V_g$ = 0 V. For *n*-type exfoliated multilayer graphene, we obtain an upper estimate of the Fermi level by using equation (1), *i.e.* $E_{Fn}$ = -4.44 eV. Thus from $E_{Fp}$ = -4.74 eV and $E_{Fn}$ < -4.44 eV, we infer a built-in electric field $E_{bi} = (E_{Fn}\text{-}E_{Fp})/el$ < 150 kV cm$^{-1}$, where $l$ = 15 nm is the thickness of the MoTe$_2$ layer. On the other hand, since exfoliated graphene is *n*-type, its Fermi level should be above the Dirac point, *i.e.* $E_{Fn}$ > -4.55 eV. Thus we conclude that the built-in electric field should be within the range 127 kV cm$^{-1}$ < $E_{bi}$ < 150 kV cm$^{-1}$. The corresponding energy band diagram of the heterostructure at equilibrium is shown in **Figure 4**(a). The junction can work under different conditions including the photovoltaic mode without any applied voltage (Figure 4(b)) and the photoconductive mode under reverse bias (Figure 4(c)). At zero bias, the built-in electric field points in same direction at the CVD graphene/MoTe$_2$ and exfoliated graphene/MoTe$_2$ interfaces. Thus the photoresponse of the regions where all layers overlap is stronger than in other regions. This is consistent with our scanning photovoltage maps (Figure 3(c-d)) and studies of additional devices including MoTe$_2$ with symmetric exfoliated graphene contacts and MoTe$_2$ with symmetric CVD graphene contacts (Figure 1(d)). The curves show nonlinear characteristics, which confirms that contact barriers exist at the interfaces of CVD graphene/MoTe$_2$ and exfoliated graphene/MoTe$_2$. Finally, we note that under a reverse bias voltage, the externally applied electric-field points in the same direction as the built-in electric field. At $V_{ds}$ = -0.2 V, the electric field almost doubles, leading to a larger photocurrent, as measured experimentally (Figure 2).

The stability and speed of the photoresponse are crucial figures of merit of a photodetector. **Figure 5**(a) and (b) show the temporal response of the photocurrent. This is obtained with a square-wave modulation of the light intensity for different powers ($P$ = 0.31, 1.25, and 3.13 W cm$^{-2}$) at $\lambda$ = 1064 nm. Under zero or reverse biases, the photocurrent can be switched on and off repeatedly and reproducibly. This switching behavior was also observed for photoexcitation under different laser wavelengths ($\lambda$ = 400, 550, 635, 800, and 1064 nm) at zero bias (**Figure S6**). Also, the heterostructure exhibits a fast dynamic response (Figure 5(c) and (d)). To study the temporal response of the current, the heterostructure was illuminated with pulsed light generated by a light-emitting diode driven by a square-wave signal generator. The dynamic response for the rise and decay of the photocurrent is well described by the equations $I(t)=I_0[1 - exp(-t/\tau_r)]$ and $I(t)=I_0 exp(-t/\tau_d)$, where $\tau_r$ and $\tau_d$ are the time constants for the rise and decay of the current. By fitting the rising and falling edges of the current versus time in Figure 5(c), we derive $\tau_r$ = 16.58 μs and $\tau_d$ = 14.96 μs at $V_{ds}$ = 0 V. A faster photoresponse with $\tau_r$ = 6.15 μs and $\tau_d$ = 4.35 μs is obtained at $V_{ds}$ = -0.20 V. This faster dynamics is assigned to the enhanced electric field of the heterostructure by the applied reverse bias.[28] The measured photoresponse times are faster than those recently reported for FETs and heterostructure photodetectors in the recent literature.[1,16,40-42] The improved photoreponse arise from the short transport channel and enhanced built-in electric field of the heterostructure.

## 3. Conclusions

In summary, we have demonstrated high-performance heterostructure photodetectors based on multilayer $MoTe_2$ with vertical asymmetric graphene contacts. The heterostructure not only exhibits a broadband photoresponse from $\lambda$ = 400 to 1400 nm, but also shows high responsivity of up to $R$ = 27.64 mA $W^{-1}$ under zero bias. Thus even without any externally applied voltage, the photo-generated carriers are efficiently separated by the built-in electric field of the heterostructure. The photoresponse is significantly enhanced in the vertical overlapping region (exfoliated graphene/$MoTe_2$/CVD graphene). Furthermore, the heterostructure shows a fast temporal response with decay and rise times in the microsecond range. The improved photoresponse indicate that van der Waals heterostructures with asymmetric graphene contacts are promising candidates for high-speed and self-powered optoelectronic devices.

## 4. Experimental sections

**Device fabrication.** The multilayer graphene was mechanically exfoliated using adhesive tape from a bulk single crystal. We find and identify the approximate thickness of the graphene flakes by optical contrast using an optical microscope. The accurate thickness of the flakes was determined by atomic force microscopy (AFM). The multilayer graphene was firstly transferred on a Si/SiO$_2$ substrate (300 nm-thick SiO$_2$) and served as bottom contact. Using the same mechanical exfoliation and transfer method, a MoTe$_2$ flake was then transferred onto the bottom graphene layer. Finally, a CVD graphene microstamp was transferred onto the MoTe$_2$ sheet to form the top contact. The CVD graphene was synthesized on a Cu substrate. By wet-etching, electron beam lithography and oxygen plasma etching, CVD-graphene was processed into microstamps prior to the mechanical transfer. Metallic contacts (Ta/Au) were fabricated on the substrate using standard photoetching, magnetron sputtering and lift off. All mechanical exfoliation and transfer processes were conducted inside a glove box.

**Electrical and opto-electrical studies.** The $I_{ds}$ - $V_{ds}$ curves were measured by an Agilent Technology B1500A Semiconductor Device Analyzer. The measurement resolution of the Semiconductor Device Analyzer was down to 10 fA and 0.5 μV. The monochromatic illumination was provided by a Zolix Omni-λ300 monochromator with a Fianium WhiteLase Supercontinuum Laser Source. A objective lens (100×, Olympus), a light chopper, a micromechanical stage (MAX381, Thorlabs) and a lock-in amplifier (SR830, Stanford

Research Systems) were used to carry out spatially resolved photovoltage mapping. A digital oscilloscope and pulsed light were used to investigate the response time. The pulsed light was generated by a light-emitting diode driven by a square-wave signal generator.

## Supporting Information

Supporting Information is available from the Wiley Online Library or from the author.

## Acknowledgments

This work was supported by National Key R&D Program of China No. 2017YFA0303400, by the NSFC Grant No. 61774144, by the EU Graphene Flagship Project, and the Engineering and Physical Sciences Research Council (Grant No. EP/M012700/1). The Project was sponsored by CAS, Grant NO. QYZDY-SSW-JSC020, XDPB12, and XDB28000000, and also by K C Wong Education Foundation.

## Conflict of Interest

The authors declare no conflict of interest.

**Figures**

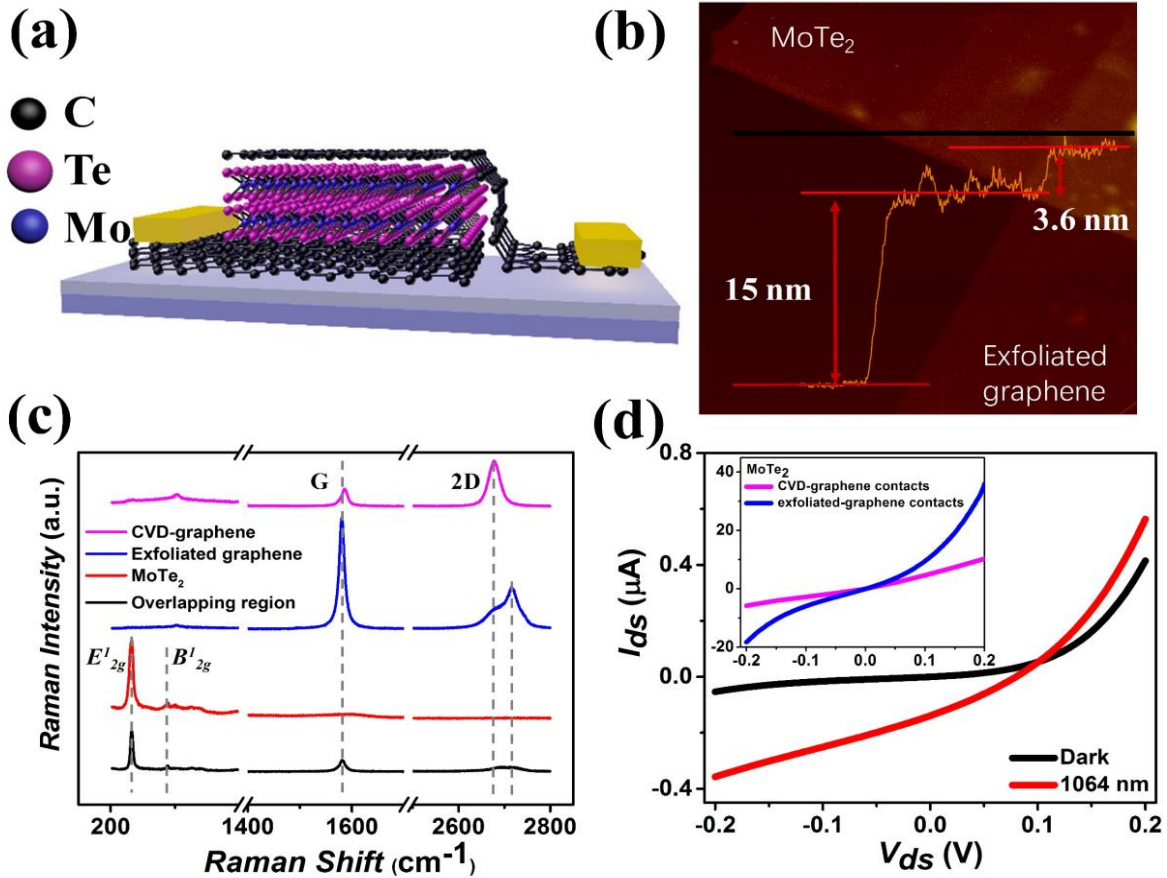

**Figure 1** (a) Schematic diagram of multilayer MoTe$_2$ with asymmetric graphene contacts. (b) AFM image of one typical device. The inset is the AFM z-profile showing the thickness of the MoTe$_2$ and bottom exfoliated graphene layers. (c) Raman spectra for different regions of the heterostructure including CVD-graphene, exfoliated graphene, MoTe$_2$, and the vertical overlapping region. The diameter of the laser spot for the Raman studies is approximately 1.5 μm. (d) Current-voltage $I_{ds} - V_{ds}$ curves in the dark and under illumination with a laser of wavelength $\lambda$ = 1064 nm and power $P$ = 3.13 W cm$^{-2}$. The diameter of the laser beam is about 45 μm, which is larger than the device size. The inset shows the $I_{ds} - V_{ds}$ curves in the dark of MoTe$_2$ photodetectors with symmetric graphene contacts.

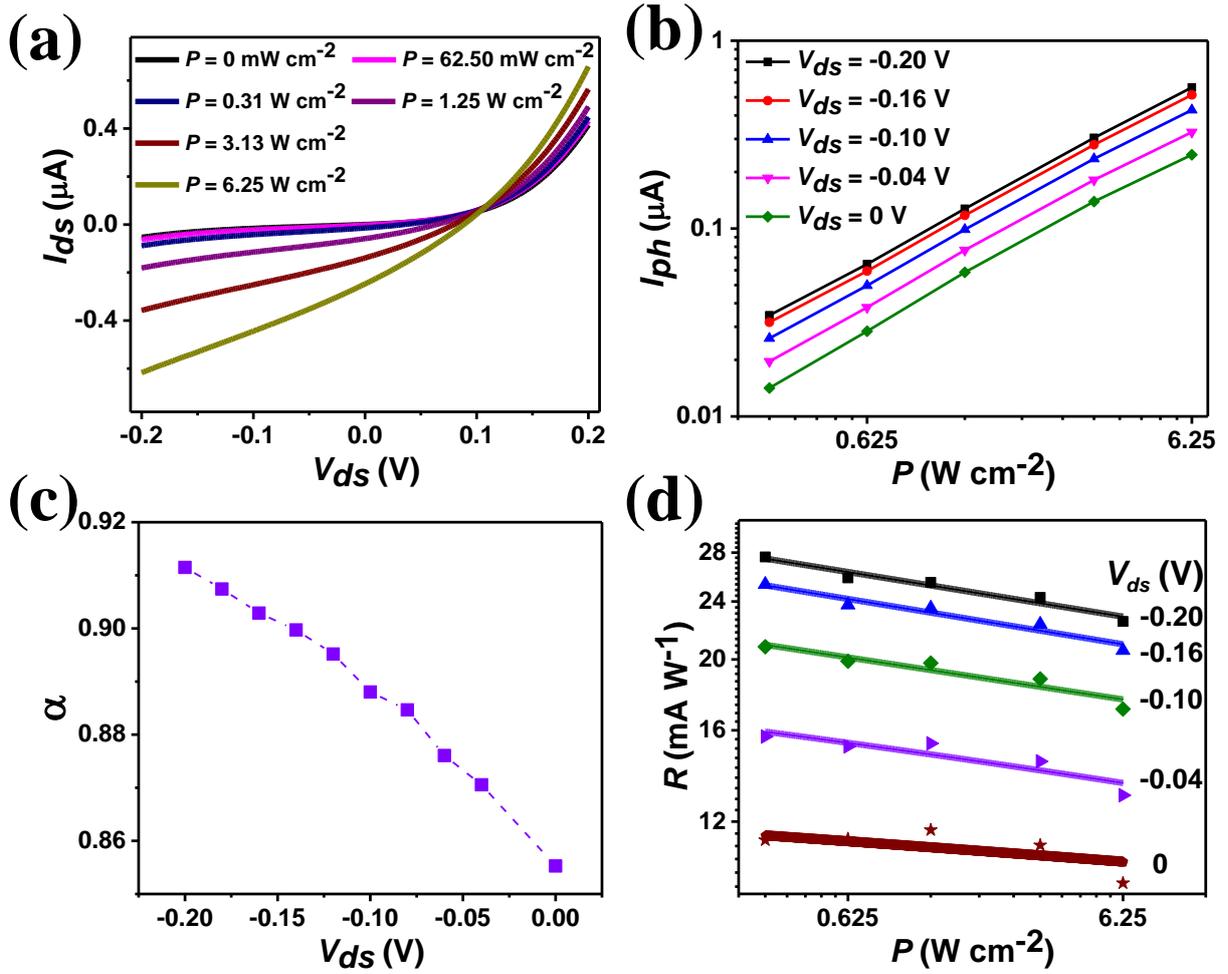

**Figure 2** (a) $I_{ds}$ - $V_{ds}$ curves with illumination at various laser powers $P$ at room temperature ($\lambda$ = 1064 nm). (b) Photocurrent $I_{ph}$ as a function of $P$ at different reverse biases $V_{ds}$. (c) Dependence of the exponent α on the reverse bias ($\lambda$ = 1064 nm). (d) Photoresponsivity $R$ as a function of $P$ at different $V_{ds}$. The diameter of the laser spot is about 45 μm, which is larger than the device size.

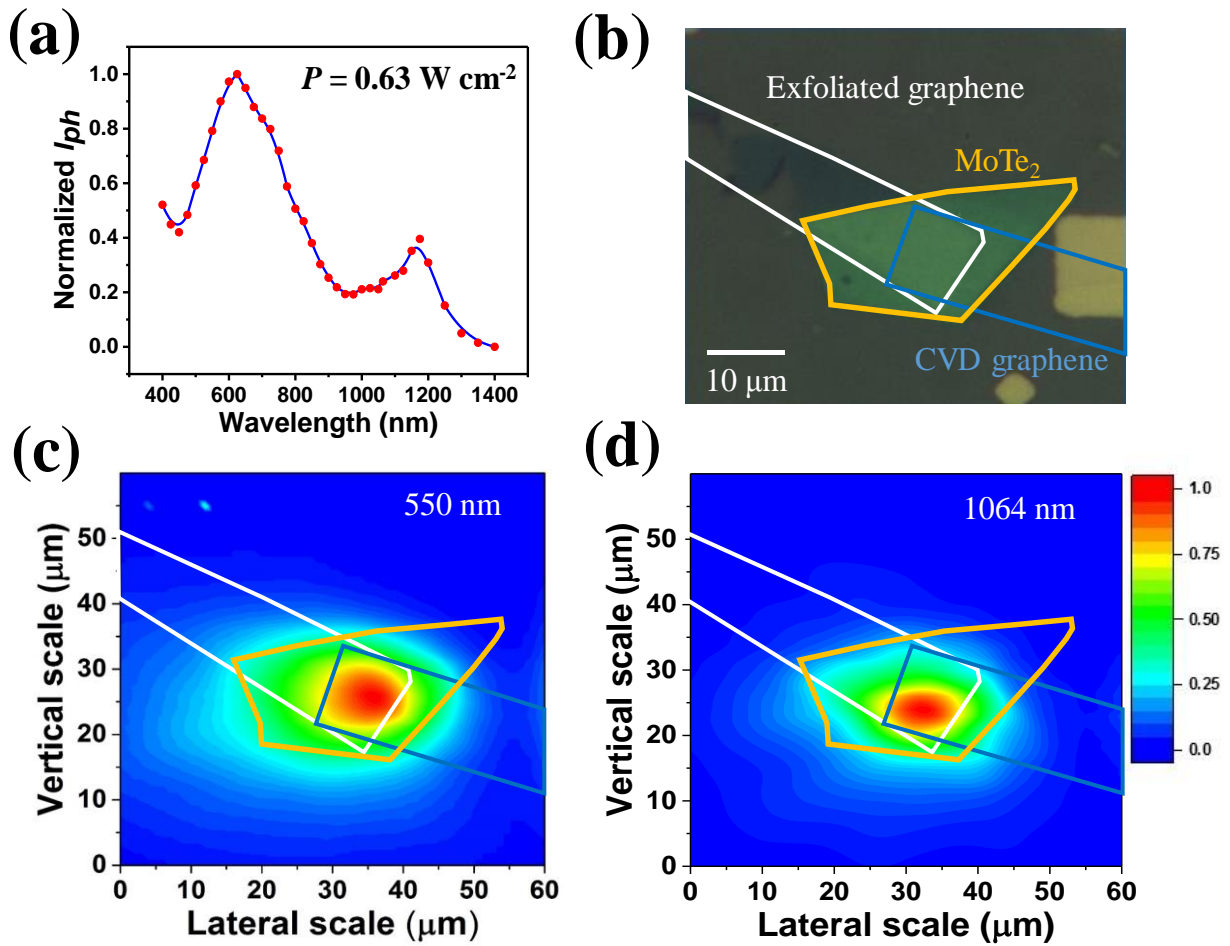

**Figure 3** (a) Normalized photocurrent $I_{ph}$ as a function of illumination wavelength from 400 to 1400 nm at zero bias, with incident power $P$ = 0.63 W cm$^{-2}$. (b) Optical microscope image of the heterostructure. (c) and (d) Normalized photovoltage maps at zero bias of multilayer MoTe$_2$ with asymmetric graphene contacts. The maps are obtained by scanning a focused laser beam with wavelength $\lambda$ = 550 nm and 1064 nm, respectively ($P$ = 20 μW). The white, yellow and blue lines demark different layers: exfoliated graphene, MoTe$_2$ and CVD graphene, respectively. The largest photoresponse is observed in the vertical overlapping region (exfoliated graphene/MoTe$_2$/CVD graphene). The laser beam is focused by a 100 × microscope objective to a spot of diameter $d$ ~ 1.5 μm.

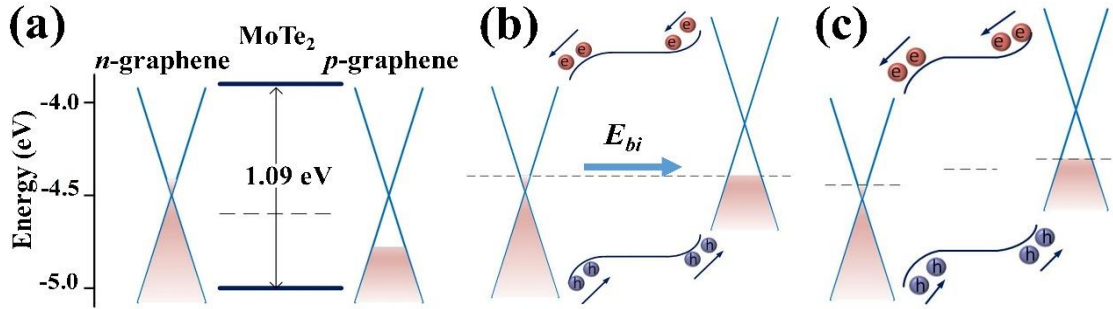

**Figure 4** (a) Energy band alignment for isolated *n*-type exfoliated graphene, *p*-type MoTe$_2$, and *p*-type CVD graphene. (b) and (c) Energy band diagrams of the heterostructure under illumination at zero bias and reverse bias, respectively. The dotted line represents the Femi level. The blue arrow shows the direction of the built-in electric field.

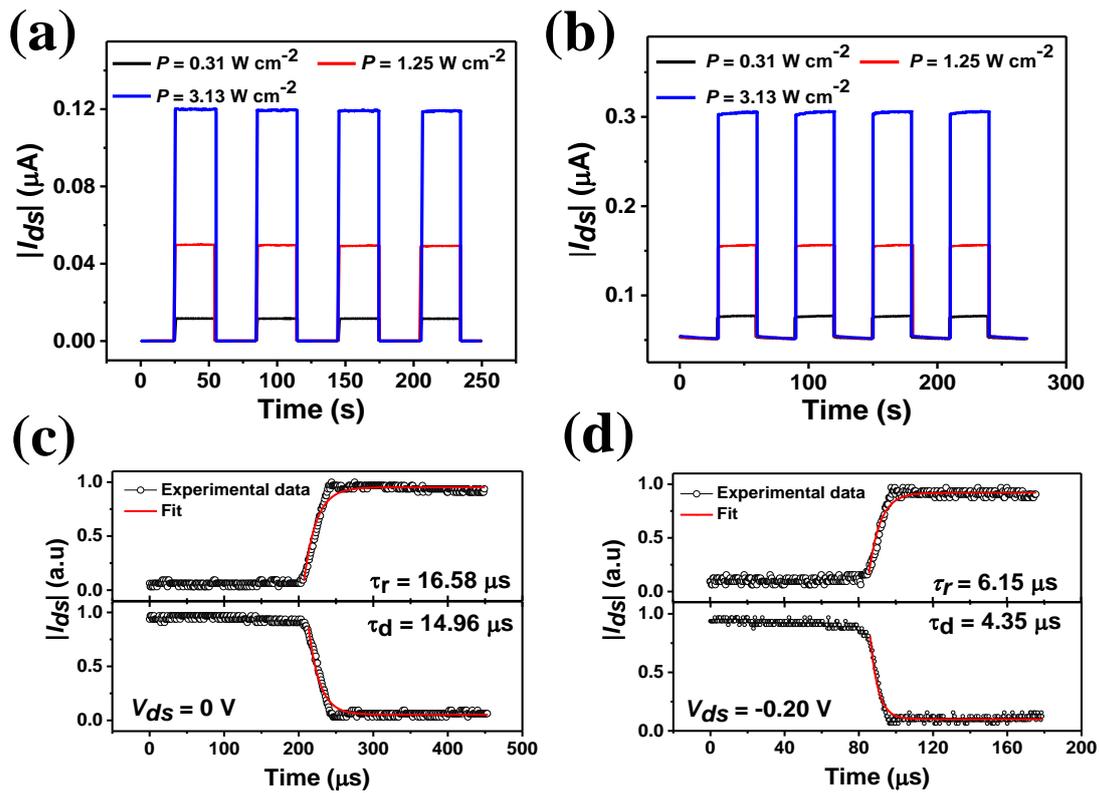

**Figure 5** (a) and (b) Reproducible photo-switching of the heterostructure with 1064 nm illumination under different powers ($P$ = 0.31, 1.25, and 3.13 W cm$^{-2}$) at $V_{ds}$ = 0 V and -0.20 V, respectively. (c) and (d) Temporal dependence of the photocurrent at $V_{ds}$ = 0 V and $V_{ds}$ = -0.20 V at room temperature. The red solid lines are fits to the data.